\documentclass[letterpaper, 12pt]{article}

\usepackage{epsfig,psfrag,amsmath,amssymb,latexsym}
\usepackage[usenames,dvipsnames]{color}
\usepackage{amscd}
\usepackage{amsfonts}
\usepackage[top=2cm, bottom=2cm, left=2cm, right=2cm]{geometry}
\usepackage{appendix}

\newcommand{\E}{\mathbf{E}}
\newcommand{\x}{\mathbf{x}}
\newcommand{\p}{\mathbf{p}}
\newcommand{\n}{\mathbf{v}}
\newcommand{\kk}{\mathbf{k}}
\newcommand{\M}{{\cal M}}
\newcommand{\J}{{\cal J}}

\newcommand{\bH}{\mathbf{H}}

\begin{document}

\title{Spatial Structure of
Stationary Nonequilibrium States in the Thermostatted Periodic Lorentz Gas.}
\date{\today}
\author{F. Bonetto,\\
\emph{\small School of Mathematics, Georgia Institute of
Technology, Atlanta GA 30332} \\
N. Chernov, A. Korepanov,\\
\emph{\small Department of Mathematics, University of Alabama at
Birmingham,
Birmingham AL 35294} \\
J.L. Lebowitz.\\
\emph{\small Departments of Mathematics and Physics, Rutgers
University, Piscataway NJ 08854} }

\maketitle

\abstract{We investigate analytically and numerically the spatial structure of
the non-equilibrium stationary states (NESS) of a point particle moving in a
two dimensional periodic Lorentz gas (Sinai Billiard). The particle is subject
to a constant external electric field $E$
as well as a Gaussian thermostat which keeps the speed $|\mathbf{v}|$ constant.
We show that despite the singular nature of the SRB measure its projections on
the space coordinates are absolutely continuous. We further show that these
projections satisfy linear response laws for small $E$. Some of them are
computed
numerically. We compare these results with those obtained from simple models in
which the collisions with the obstacles are replaced by random collisions.
Similarities and differences are noted.}

\section{Introduction}\label{intro}

In this paper we continue our study of nonequilibrium stationary states
(NESS) maintained by a Gaussian thermostat \cite{BDL,CELS,BL,Ch}.
Theoretical analysis and computer simulations show that the NESS
obtained from such artificial model dynamics can give useful
information on real systems maintained in NESS by coupling with heat baths
\cite{Therm}.

Here we focus on the Moran-Hoover (MH) model of a single
particle in a periodic billiard moving under the influence of an electric
field $\E$ and a Gaussian thermostat that keeps the kinetic energy
constant \cite{MH}. The equations of motion are:

\begin{equation}\label{motion}
\left\{
 \begin{array}{l}
  \dot \x= \p\crcr
  \dot \p= \E -\alpha(\p)\p+\mathbf{F}_{\rm obs}(\x)\crcr
  \alpha(\p)=\frac{(\p\cdot \E)}{(\p\cdot \p)}
 \end{array}\right.
\end{equation}
where $\x$ is the position, $\p$ the momentum of the particle with
unit mass, and $\mathbf{F}_{\rm obs}(\x)$ represents elastic collision with the
obstacles. It
is clear from eq.(\ref{motion}) and the fact that collisions with the
obstacles do not change $|\p|$ that $\frac{d}{dt}(\p\cdot \p)=0$. We shall
therefore set $|\p|=1$ from now on.

\begin{figure}[ht]
  \centering
     \epsfig{file=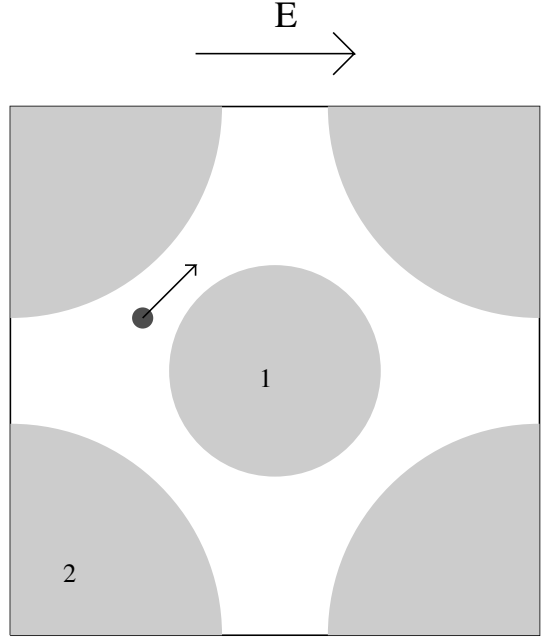,width=0.4\linewidth}
   \centering
 \caption{Typical obstacles placement.}
 \label{figure1}
\end{figure}

The particle moves on a 2-dimensional torus whose side can be
chosen to be 1. An arrangement of the  obstacles used for all
the simulations presented in this paper is shown in
Figure \ref{figure1}. The two obstacles have radii $r_1=0.2$ and $r_2=0.4$. This
is also
the arrangement used in our previous works \cite{BDL,BDLR} and is chosen to
have a finite horizon, {\sl i.e.} there is an upper bound for the time between
successive collisions of the particle with the obstacles. Moreover we
take $\E$ to be along the horizontal $x$ axis, {\sl i.e.} $\E=(E,0)$. The
analytical results apply to general geometries with finite horizons.

\begin{figure}
 \centering
   \epsfig{file=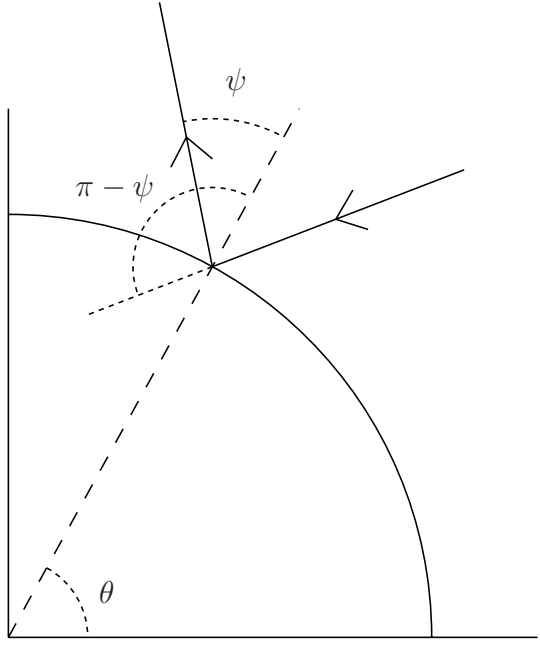,width=0.4\linewidth}
   \put(-110,210){\makebox(0,0)[r]{$\psi$}}
   \put(-145,170){\makebox(0,0)[r]{$\pi-\psi$}}
   \put(-160,17){\makebox(0,0)[r]{$\theta$}}
  \centering
\caption{Elastic collision.}
 \label{figure2}
\end{figure}

The set of states where the particle collides with a given obstacle can
be parametrized by two angles: $\vartheta\in[0,2\pi]$ the angle on the
obstacle between the collision point and the positive $x$ direction,
and $\psi$ the angle between the particle velocity and the outgoing
normal to the obstacle at the collision point. To obtain a complete
coordinate system for the collision states we define the coordinate
$\theta=\vartheta$ for obstacle 1 (see Figure \ref{figure1}) and
$\theta=\vartheta+2\pi$ for obstacle 2 so that $\theta\in[0,4\pi]$. In
these coordinates, the elastic collision is simply represented by the
map $C(\theta,\psi)=(\theta,\pi-\psi)$, where $\psi\in[\pi/2,3\pi/2]$
before collision and $\psi\in[-\pi/2,\pi/2]$ after collision. We will
call $\M=[0,4\pi]\times [-\pi/2,\pi/2]$ the set of possible pairs
$(\theta,\psi)$ representing the position of the particle just after a
collision. $\M$ corresponds to a Poincar\'e section of the flow. See
Figure \ref{figure2} for a depiction of $\theta$ and $\psi$.

Since $|\p|$ is constant, the trajectory of the
particle can be represented by its position $\x(t)$ and the angle of its
momentum with the horizontal axis $\phi(t)$. The motion of the particle between
two collisions can be exactly integrated. Moreover one can construct the map
$S_E(\theta,\psi)$ mapping the position and momentum of the particle just after
a collision to its position and momentum just before the next collision, which
may be with the same or a different obstacle.

In this way we can represent the dynamics in discrete time as the
iteration of the map $T_E:\M\to\M$ between successive collisions given
by $T_E=C\circ S_E$.
Observe that this map is not continuous (grazing collisions) and that,
for $E$ small, the dynamics is a perturbation of the free billiard
dynamics \cite{CELS}.

In our previous works we were primarily concerned with the SRB
distribution associated with $T_E$. Let $\mu_0$ be the
measure on $\M$ given by 
\begin{equation}\label{mu0}
\mu_0=\cos(\psi)\chi(\theta)d\theta\,d\psi/Z
\end{equation}
where
$Z=4\pi(r_1+r_2)$ is a normalization constant and $\chi(\theta)=r_1$ for
$\theta\in[0,2\pi]$
and $\chi(\theta)=r_2$ otherwise. Observe that $\mu_0$ is invariant under
$T_0$. The SRB distribution $\mu_E(d\theta,d\psi)$ is
defined as the weak limit of $\mu_0$ under the dynamics $T_E$,
{\sl i.e.}

\begin{equation}\label{wlim}
 \mu_{E}=\operatorname{w-lim}\, T_{E}^n \mu_0.
\end{equation}
The measure $\mu_E$, when it exists and is
unique, represents the natural non equilibrium steady state (NESS) for the
system \cite{Ru}. Clearly $\mu_0$ is the SRB measure of $T_0$.

From the SRB measure $\mu_{E}(d\theta,d\psi)$ on $\M$ for the collision
map one can build the SRB measure $m_{E}(d\x,d\phi)$ on $M=Q\times
[0,2\pi]$ for the flow generated by eq.(\ref{motion}). Here $Q$ is
$\mathbb{T}\backslash$~obstacles. This can be represented as:

\begin{equation}\label{mmu}
 m_{E}(A)=
\frac{1}{\bar{\tau}_E}\,\int_\M\int_0^{\tau_E(\theta,\psi)}
I_A(\mathbf{X}^E_t(\theta,\psi),\Phi^E_t(\theta,\psi))\,dt\,
\mu_E(d\theta,d\psi)
\end{equation}
where $I_A$ is
the indicator function of the set $A\subset M$,
$(\mathbf{X}^E_t,\Phi^E_t)$ is the flow generated by eq.(\ref{motion}) and
$\tau_E(\theta,\psi)$ is the time till the next collision when starting
at $(\theta,\psi)\in\M$ with  $\bar{\tau}_E = \int_{\M}
\tau_E(\theta,\psi)\,\mu_E(d\theta,d\psi)$ denoting the mean free time.

In \cite{CELS} it was proved that, for small fields $E$, $|E|<E_0$, the above
model has a unique NESS described by an SRB measure $\mu_E$ which is singular
with respect to the Liouville measure with Hausdorff dimension given by the
standard Kaplan-Yorke formula \cite{FKYY}.

The current $\mathbf{j}(E)$ in this NESS is given by
\[
 \mathbf{j}(E)=m_E(\mathbf{v})
\]
where $\mathbf{v}=(\cos(\phi),\sin(\phi))$ is the velocity of the
particle. This current was shown in \cite{CELS} to be given by the
Kawasaki formula cf. \cite{KE}. In the limit $E\to 0$ the Kawasaki formula
reduces
to the Green-Kubo formula for the conductivity $\kappa$ which satisfies
the Einstein relation \cite{CELS,KE}. An investigation of the current
as a function of the field was carried out in \cite{BDL}. It was argued
there that the current is not a $C^1$ function of the field $E$ even
close to $E=0$. The results of \cite{CELS} were generalized in
\cite{Ch,Z} to systems where the collision rule or the free flow
dynamics is perturbed.

In none of the above works was the spatial dependence of the singular (with
respect to Lebesgue)
measure $m_E(d\x,d\phi)$ studied. This is what we do in this note. We will
describe analytical results and numerical studies of the spacial and
angular dependence of the NESS $m_E(d\x,d\phi)$ when projected on
$\phi\in[0,2\pi]$ or
on $\x\in Q$ and related quantities like the local average velocity.

The outline of the rest of the paper is as follows. In section \ref{s:avcu} we
introduce the local density, local average velocity and angular distribution
derived from $m_E$. We find their dependence on position and
field strength. We also show there
computer generated pictures of the flow and compare them with the predictions
of Green-Kubo formulas at small $E$. In section
\ref{s:stocha} we introduce and analyze two simple models in which the
deterministic collisions with fixed obstacles are replaced by random collisions
whose times form a Poisson process and compare their properties with those of
the deterministic model. The appendixes are devoted to analytical justification
of the claims in section \ref{s:avcu}. A paper describing results for the case
where the system consists not just of one but of a large number of particles is
in preparation.

\section{Local Structure of the SRB measure}\label{s:avcu}

We define and study the several projections of the SRB measure
$m_E$ introduced in the previous section. For clarity of exposition we delay
derivations and justifications to the Appendices.

\subsection{Local Density and Average Velocity}\label{ss:2.1}

Two interesting quantities to study are {\sl the local density} and
{\sl local average velocity}. More precisely, we define the projected measures
on the position $\x$ as:

\begin{equation}\label{dd}
\delta_E(A)=\int_{A\times[0,2\pi)} m_{E}(d\x,d\phi)
\end{equation}
for any $A\subset Q$. This clearly defines a
probability measure $\delta_E(d\x)$ on $Q$. Using eq.(\ref{mmu}), and
defining

\begin{equation}\label{JA}
 J_A^E(\theta,\psi)=\frac{1}{\bar\tau_E}
\int_0^{\tau(\theta,\psi)}
I_{A}\left(\mathbf{X}^E_t(\theta ,
\psi)\right)\,dt
\end{equation}
where $I_A$ is the indicator function of the set $A$, we can
represent $\delta_E(A)$ as the integral of a piecewise smooth function
with respect to $\mu_{E}(d\theta,d\phi)$:
\begin{equation}  \label{deltaA}
 \delta_E(A)=\int_\M  J_A^E(\theta,\psi)\, \mu_{E}(d\theta,d\psi)
\end{equation}
Observe that $J_A^E(\theta,\psi)$ is the relative amount of time the trajectory
starting from $(\theta,\psi)$ spends in the set $A$, that is the amount of time
divided by the mean free time $\bar\tau_E$.

We also define the vector measure for the local average velocity

\begin{equation}\label{nunu}
\nu_E(A)=\int_{A\times[0,2\pi)}(\cos(\phi),\sin(\phi))\, m_E(d\x,d\phi)
\end{equation}
Also this measure can be written as the integral of a piecewise smooth function
with respect to $\mu_{E}(d\theta,d\phi)$, see eq.(\ref{HH}) below for details.

In Appendix \ref{a:regu} we show that, for $|E|<E_0$, the integrals
in eq.(\ref{dd}) and eq.(\ref{nunu}) define absolutely continuous
measures. That is
$\delta_E(d\x)=n_E(\x)\,d\x$ and $\nu_E(d\x)=n_E(\x)\n_E(\x)\,d\x$. We
call $n_E(\x)$ the {\sl local density} and $\n_E(\x)$ the {\sl local
average velocity} at $\x$. We show that both are continuous functions
of both $\x$ and $E$ with $n_0(\x)=$const$=\bigl({\rm
Area}(Q)\bigr)^{-1}$ and $\n_0(\x)=0$.

\begin{figure}[ht]
  \centering
     \epsfig{file=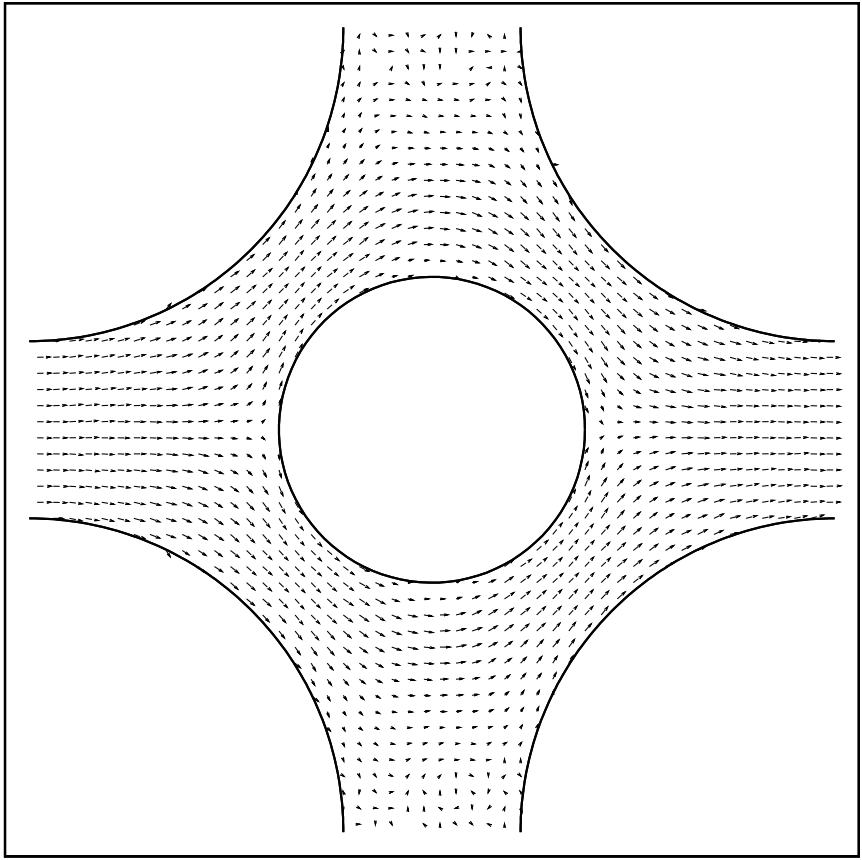,width=0.7\linewidth,clip=}
  \caption{Average local velocity $\n_E(\x)$ for $E=0.1$.}
 \label{Figure_Velo}
\end{figure}

To visualize the above numerically, we divided the torus of Figure~\ref{figure1}
in a grid of $50\times 50$ cells and computed the time
average of the velocity of the particle when it crosses a cell. The
results are shown in Figure \ref{Figure_Velo}. We also computed the local
density on the same grid.

We now show that the local density $n_E(x)$ and the local
average velocity $\n_E(\x)$ are linear in the field $E$ when $E\to0$, that is
\begin{eqnarray}
n_E(\x)&=&n_0(\x)+d(\x)E+o(E)\label{dx}\\
\n_E(\x)&=&\kk(\x)E+ o(E)\label{nx}
\end{eqnarray}
where $d(\x)$ and $\kk(\x)$ can be computed via Green-Kubo-type formulas as
follows.

Consider the family of all velocity vectors originating at the point
$\x$ (at which we are computing the density or average velocity); they
make a one parameter family of phase states  $W_{\x} = \{(\x,\phi)\ |\
0< \phi<2\pi\}$. Let $\rho^{\x}$ be the probability measure on $W_{\x}$
that has a uniform distribution over $\phi\in[0,2\pi]$. We can map
$W_{\x}$ to the collision space $\M$ by taking every point
$(\x,\phi)\in W_{\x}$ to its first collision with $\partial Q$ in
\emph{the past}, under the field-free dynamics.
The image of $W_{\x}$ will then be a collection $W_0$ of curves in $\M$
on which we get an induced probability measure $\rho_0$. Pulling this
measure further back (into the past) we get a sequence of probability
measures $\rho_n = T_0^{-n}(\rho_0)$, each sitting on a collection
$W_n=T_0^{-n}(W_0)$ of curves in $\M$. With this definition we get that

\begin{equation}\label{d0}
 d(\x)=
c\biggl[\rho_0(\Delta_{0,\x})+\sum_{n=1}^{\infty}
  \rho_{n}(\Delta_0)\biggr]
\end{equation}
where $c={\rm Area}(Q)^{-1} = n_0(\x)$,
$\Delta_0=\tau_0(\theta,\psi)\cos(\theta+\psi)$ is the $x$-distance
form the collision point $(\theta,\psi)\in\M$ to the next collision
point and $\Delta_{0,\x}$ is the $x$-distance from the collision point
$(\theta,\psi)\in\M$ to the point $\x$. Observe that $\rho_0$ is
supported on points whose trajectory passes through $\x$ before
colliding again. The above series converges exponentially, because the
measures $\rho_n$ converge exponentially fast to the measure $\mu_0$ on
$\M$ (see Theorem 7.31 in \cite{CM}) and $\mu_0(\Delta_0)=0$.

Consider now the two {\it signed} measures $\rho^{c,\x}$ and $\rho^{s,\x}$ on
$W_{\x}$ that have densities $\cos\phi$ and $\sin \phi$, respectively, with
respect to $\rho^{\x}$. As before, we can map $\rho^{c,\x}$ and $\rho^{s,\x}$
on the collision space $\M$ and obtain
signed measures $\rho^c_0$ and $\rho^s_0$ on $W_0$, respectively. We also denote
their images by $\rho^{c,s}_n = T_0^{-n}(\rho^{c,s}_0)$ on $W_n$ for $n\in
{\mathbb Z}$.

\begin{equation} \label{kk}
  \kk(\x) = \tfrac{1}{2} \, \sum_{n=-\infty}^{\infty}
  \bigl(\rho^c_n(\Delta_0),\rho^s_n(\Delta_0)\bigr).
\end{equation}
The terms in the series in eq.\eqref{kk}
converge to zero as $n\to\pm\infty$ exponentially fast, because the
measures $\rho^{c,s}_n$ converge to the zero measure; this again
follows from Theorem 7.31 in \cite{CM}.

We note that the perturbation of the density $n_E(\x)$ and of the local average
velocity $\n_E(\x)$ are linear in $E$, to the leading order, and the factor of
$E$ is given, in both cases, by an infinite sum of correlations, {\sl i.e.} the
right hand sides of eq.(\ref{d0}) and eq.(\ref{kk}).

\begin{figure}[ht]
  \centering
     \epsfig{file=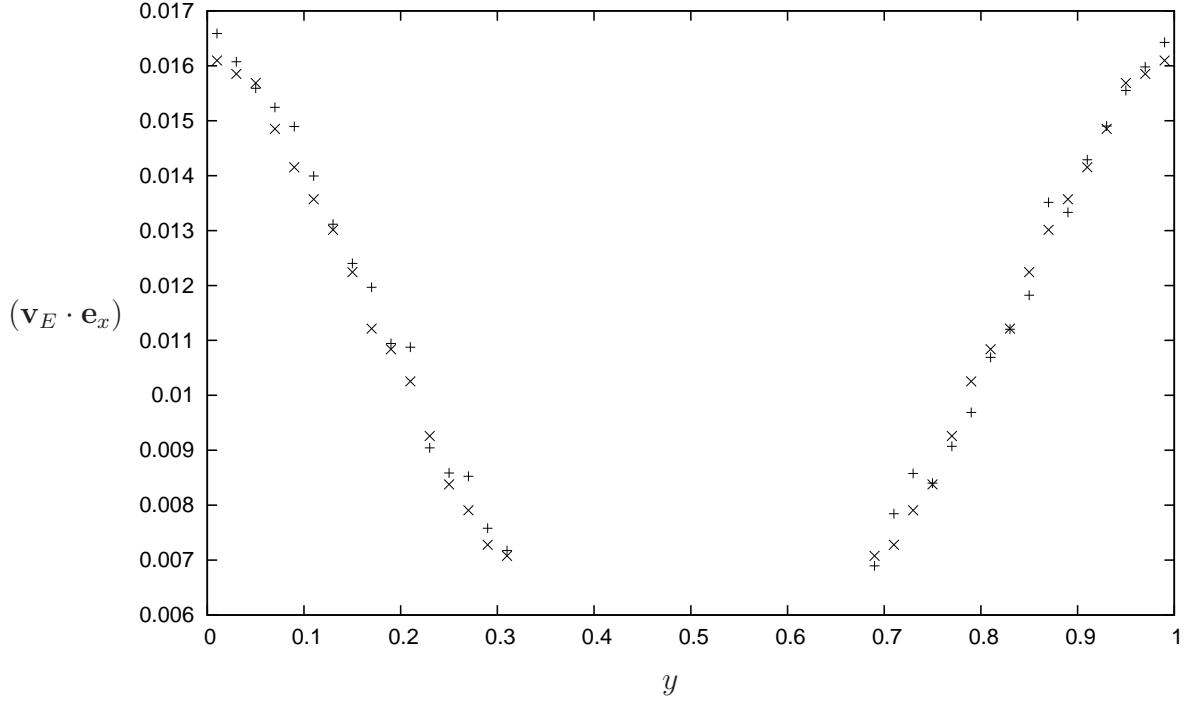,width=0.85\linewidth,clip=}
      \put(-200,-10){\makebox(0,0)[r]{$y$}}
      \put(-410,130){\makebox(0,0)[r]{$(\n_E\cdot \mathbf{e}_x)$}}
  \caption{The $x$ component of the average local velocity $\n_E(\x)$ for
$E=0.1$ and $x=0.41$.}
 \label{Figure_vx}
\end{figure}

\begin{figure}[ht]
  \centering
     \epsfig{file=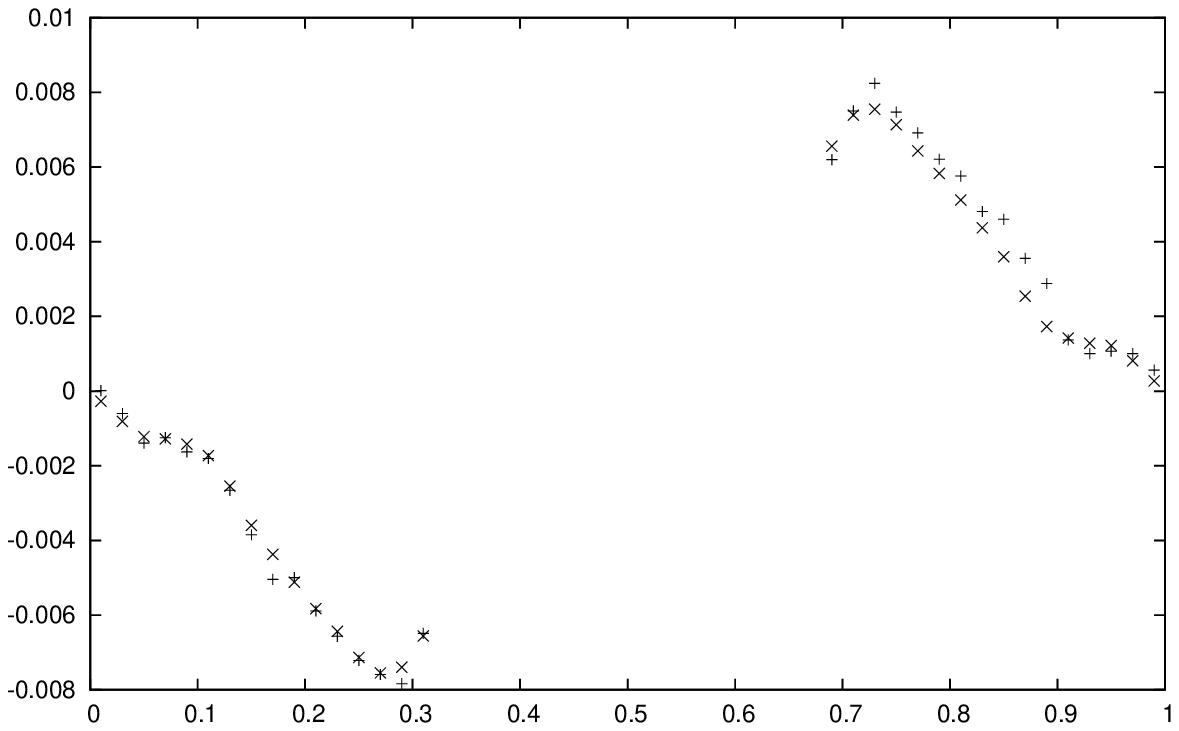,width=0.85\linewidth,clip=}
      \put(-200,-10){\makebox(0,0)[r]{$x$}}
      \put(-410,130){\makebox(0,0)[r]{$(\n_E\cdot \mathbf{e}_y)$}}
  \caption{The $y$ component of the average local velocity $\n_E(\x)$ for
$E=0.1$ and $y=0.41$.}
 \label{Figure_vy}
\end{figure}

We computed numerically the coefficients $d$ and $\kk$ in eq.(\ref{dx})
and eq.(\ref{nx}) to compare their predictions with the simulation
results shown in Figure \ref{Figure_Velo}. We truncated the infinite
sums in eqs.(\ref{d0},\ref{kk}) to $|n|<15$ since we saw no visible
difference arise from taking more terms into consideration. Let
$l^+_x=\{(x,y)\in Q\}$ be the vertical cross section  placed at horizontal
coordinate $x$ and $l^-_y=\{(x,y)\in
Q\}$ be the horizontal cross section placed
at vertical coordinate $y$. Finally let $\mathbf{e}_x=(1,0)$ and
$\mathbf{e}_y=(0,1)$ be the unit vectors in the horizontal and vertical
direction respectively. Figure \ref{Figure_vx} shows a comparison
of the horizontal component $(\n_E(\x)\cdot \mathbf{e}_x)$ of $\n_E(\x)$ along
$l^+_{0.41}$ with the
prediction of eq.(\ref{nx}). In the same way, Figure \ref{Figure_vy}
shows a comparison of the vertical component $(\n_E(\x)\cdot
\mathbf{e}_y)$ of $\n_E(\x)$ along
$l^-_{0.41}$ again with the prediction of eq.(\ref{nx}). In both
figures the pluses represent the results of direct simulation while the
crosses are obtained using the Green-Kubo formula eq.(\ref{kk}).

\begin{figure}[ht]
  \centering
     \epsfig{file=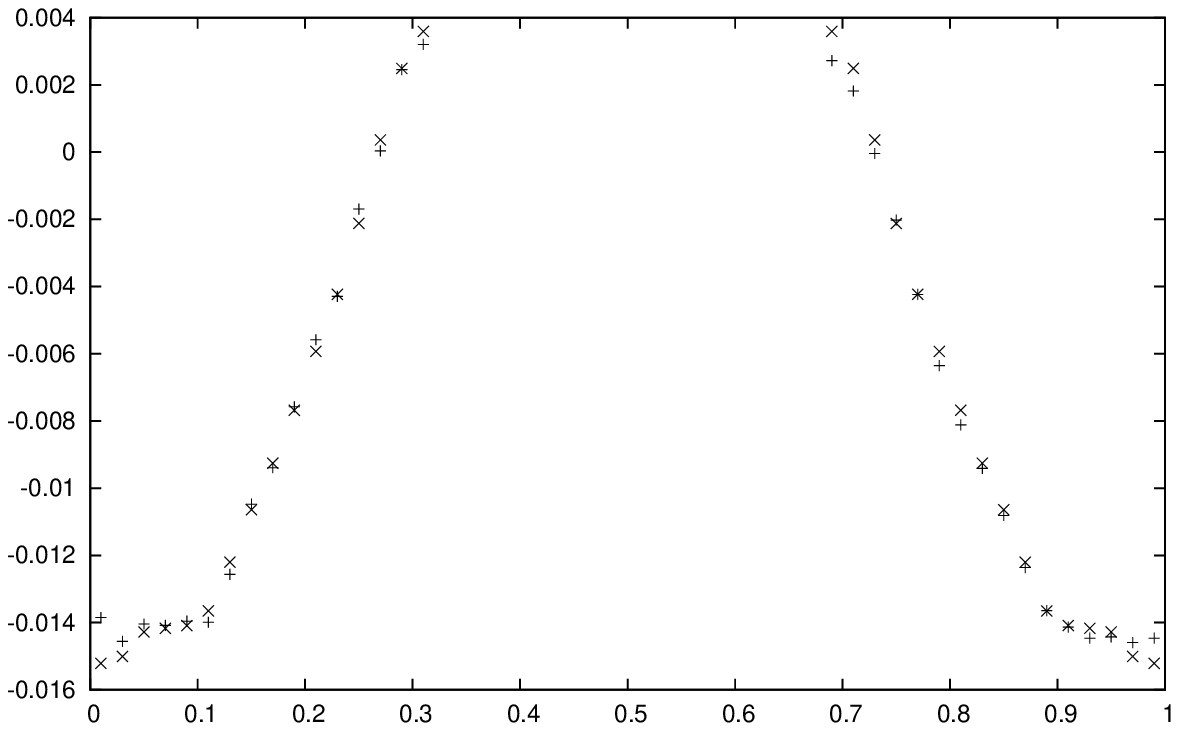,width=0.85\linewidth,clip=}
          \put(-200,-10){\makebox(0,0)[r]{$y$}}
      \put(-415,130){\makebox(0,0)[r]{$n^o_E$}}
  \caption{The symmetrized local density $n^o_E(\x)$ for $E=0.1$ and
$x=0.41$.}
 \label{Figure_d}
\end{figure}

The comparison of $n_E(\x)$ with eq.(\ref{dx}) is more difficult.
Calling $n_E^o(\x)=(n_E(\x)-n_{-E}(\x))/2$ and
$n_E^e(\x)=(n_E(\x)+n_{-E}(\x))/2-n_0(\x)$, we have that $n_E^o(\x)$
satisfies the same linear response formula eq.(\ref{dx}) of $n_E(\x)$
with the same coefficient $d(\x)$ but we expect the remainder to be
smaller. This is relevant in the present case since $n_E^e(\x)$ and
$n_E^o(\x)$ appear to be of comparable magnitude. We observe that, due
to the symmetry of the problem, $n_E(1-x,y)=n_{-E}(x,y)$ so that
$n^o_E(x,y)=(n_E(x,y)-n_E(1-x,y))/2$. Figure \ref{Figure_d} compares
$n^o_E(\x)$ along $l^+_{0.41}$ with eq.(\ref{dx}). Again the pluses
represents direct simulation while the crosses are obtained using the
Green-Kubo formula eq.(\ref{d0}).

More generally, given a probability measure $\rho(d\x,d\phi)=l(\x,\phi)\,d\x\,
d\phi$ absolutely continuous with respect to the Lebesgue measure on
$Q$ let $\rho^E_t(d\x,d\phi)=l^E_t(\x,\phi)\,d\x\, d\phi$ be its time evolution
with
respect to the dynamics generated by eq.(\ref{motion}). In a similar
way as above, we can then define:

\begin{eqnarray}
n^E_t(\x)&=&\int l^E_t(\x,\phi)\,d\phi\crcr
n^E_t(\x)\n^E_t(\x) &=&\int (\cos(\phi),\sin(\phi)) l^E_t(\x,\phi)\,d\phi
\end{eqnarray}
The density $n^E_t(\x)$ clearly satisfies a conservation law:

\begin{equation} \label{nn}
 \frac{d}{dt}\int_A n^E_t(x)\,dx=-\int_{\partial
A}n^E_t(\x)\bigl(\n^E_t(\x)\cdot \hat {\bf n}(\x)\bigr)\,d\sigma(\x)
\end{equation}
where $A$ is a subset of $Q$ with smooth enough boundary, $\hat{\bf
n}(\x)$ is the unit outward normal to $\partial A$ at
$\x$ and $\sigma(\x)$ is the length element on $\partial A$.

Taking the limit $t\to\infty$ and assuming that
$\lim_{t\to\infty}n^E_t(\x)=n_E(\x)$ and $\lim_{t\to\infty}\n^E_t(x)=\n_E(\x)$
we obtain

\begin{equation}\label{nn1}
 \int_{\partial A}n_E(\x)\bigl(\n_E(\x)\cdot \hat{\bf n}(\x)\bigr)d\sigma(\x)=0
\end{equation}

The above assumption is not trivial. It is easy to show that, if
$\lim_{t\to\infty}n^E_t(\x)$ exists, it has to equal $n_E(\x)$. On the
other hand, we do not have a proof for the existence of such a limit.
A similar argument holds for $\n^E_t(\x)$. A complete justification of
eq.(\ref{nn1}) will thus require further work but we certainly expect it to be
true.

Nonetheless we can test the validity of eq.(\ref{nn1}) numerically. Due to the
symmetry of $Q$ we
have that the average current $\mathbf{j}(E)=(j(E),0)$. Moreover, since the
collision are elastic,
$\n_E(\x)$ is tangent to $\partial Q$ for $\x\in\partial Q$. It follows from
this that

\begin{eqnarray*}
 \int_{l^+_x}n_E(\x)\bigl(\n_E(\x)\cdot \mathbf{e}_x\bigr)\,dy&\equiv&
j(E)\crcr
\int_{l^+_y}n_E(\x)\bigl(\n_E(\x)\cdot \mathbf{e}_y\bigr)\,dx&\equiv& 0
\end{eqnarray*}
independently on the value of $x$ or $y$. Both these equations are very
well verified.

\subsection{Angular Distribution}\label{s:angular}

We now look at the projection of $m_E$ on the angle $\phi$. We can
define the projected measure $\eta(d\phi)$ by setting, for a
measurable set $A\subset[0,2\pi]$,

\[
 \eta_E(A)=\int_M I_{A}(\x,\phi)\, m_E(d\x,d\phi)
\]
where $I_{A}$
is the indicator function of the set $A$. Again we can write
$\eta_E(A)$ as an integral on the SRB measure $\mu_E(d\theta,d\psi)$ as
follows. Define the function:

\begin{equation}\label{tau}
 J^E_A(\theta,\psi)=\frac{1}{\bar\tau_E}
\int_0^{\tau_E(\theta,\psi)}
I_{A}\left(\Phi^E_t(\theta ,
\psi)\right)\,dt
\end{equation}
Then we have that

\[
 \eta_E(A)=\int_\M J^E_A(\theta,\psi)\, \mu_E(d\theta,d\phi)=\mu_E(J^E_A)
\]
Using the argument in Appendix
\ref{a:regu} we can show that, for $|E|<E_0$, $\eta_E$ is absolutely
continuous with respect to $d\phi$, {\sl i.e.} that
$\eta_E(d\phi)=h_E(\phi)\,d\phi$ where $h_E(\phi)$ is a continuous
function of both $\phi$ and $E$ with $h_0(\phi)=$const$=1/2\pi$, since
the invariant measure $m_0$ is uniform on $Q\times[0,2\pi]$.

\begin{figure}[ht]
 \centering
   \epsfig{file=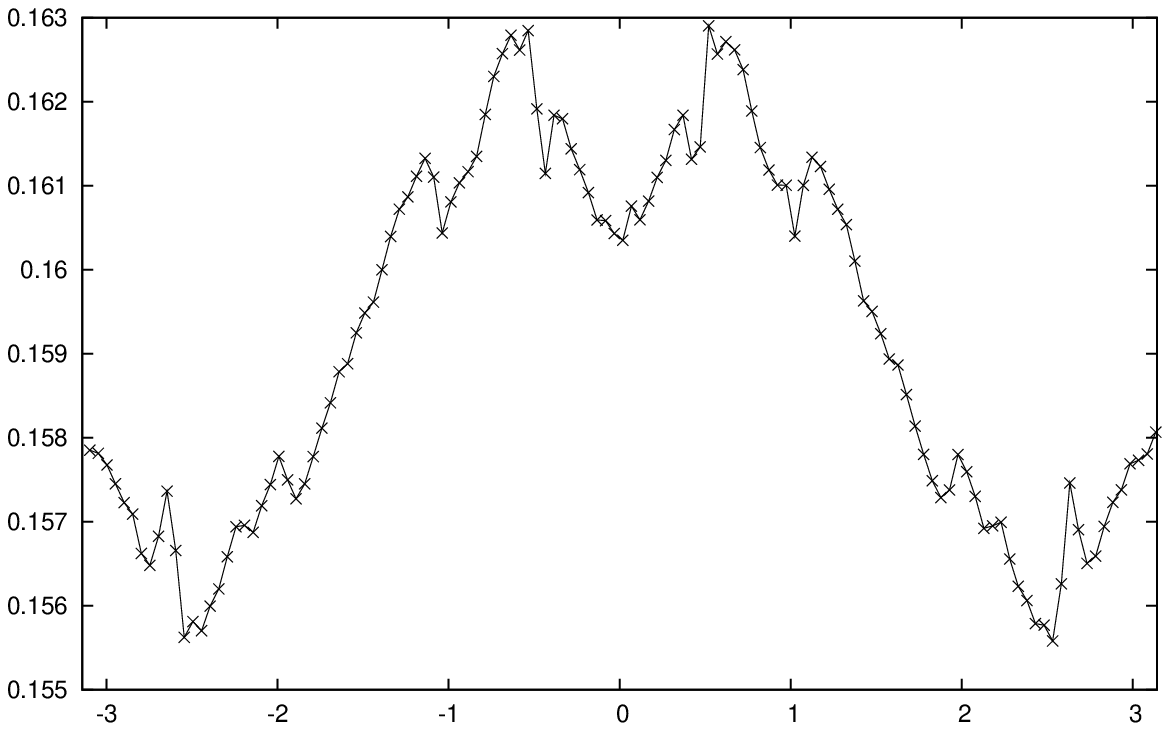, width=0.85\linewidth}
        \put(-200,-10){\makebox(0,0)[r]{$\phi$}}
      \put(-410,130){\makebox(0,0)[r]{$h_E$}}
  \centering
\caption{Angular distribution for $E=0.1$.}
\label{Figure_angle}
\end{figure}

We computed $h_E(\phi)$ numerically for $E=0.1$. The result is shown in Figure
\ref{Figure_angle}. A striking characteristic of this distribution is
the dip around $\phi=0$. This is somewhat unexpected since the effect
of the field $E$ is to push the velocity of the particle to align with
the positive $x$ direction so that one would expect a maximum at
$\phi=0$ rather than a local minimum (see also section \ref{s:stocha}
for a comparison with the stochastic models).

To understand this better we consider, for a given $\phi$, all
points $(\theta,\psi)\in\M$ that produce the outgoing velocity vector
$(\cos\phi, \sin\phi)$, i.e., we consider
$$
   V_{\phi} = \{(\theta,\psi)\in\M \colon \psi + \theta = \phi
   \ \ ({\rm mod}\ 2\pi)\}.
$$
Now $\M$ is foliated by the lines $\{V_{\phi}\}$, $0\leq \phi <2\pi$.
Let $\mu_{0}^{\phi}$ denote the conditional measure induced by $\mu_0$
on the line $V_{\phi}$. If we use $\theta$ as the (only) coordinate on
$V_{\phi}$, then
$$
   d\mu_0^{\phi} = Z_{\phi}^{-1} \cos(\phi-\theta) \chi(\theta)\, d\theta
$$
where $Z_{\phi}$ is the normalizing factor
\begin{equation} \label{Zphi}
 Z_{\phi} = \int_{\cos(\phi-\theta)>0} \cos(\phi-\theta) \chi(\theta)\, d\theta
= 2(r_1+r_2).
\end{equation}
We remind the reader that $\chi(\theta)=r_1$ and $0\leq \theta< 2\pi$
on the first obstacle and $\chi(\theta)=r_2$ and $2\pi\leq \theta<
4\pi$ on the second.

Now we consider the conditional distribution of the free flight
time function $\tau_0$ on each line $V_{\phi}$. It turns out that its
first moment is constant, i.e., $\mu_0^{\phi}(\tau_0) = \bar{\tau}_0$ for
all $\phi$'s, where $\bar{\tau}_0 = \mu_0(\tau)$ is the total
(unconditional) mean free time. In other words, the deterministic
collision process is isotropic, on average. This seems to be a novel
result in the studies of billiards and we prove it in Appendix
\ref{a:tau}. We now argue that the observed dip near $\phi=0$,
for small $E$ can be traced to the second moment,
$\mu_0^{\phi}(\tau_0^2)$, which is \emph{not} constant and which for our
obstacles indeed has a local minimum at $\phi=0$.

We will show that the density $h_E(\phi)$ satisfies
\begin{equation}\label{ax}
 h_E(\phi)=\frac{1}{2\pi}+a(\phi)E+o(E)
\end{equation}
where $a(\phi)$ is given by a Green-Kubo formula
\begin{equation}\label{kax}
 a(\phi)=\frac{Z_\phi}{Z}\frac{1}{2\bar{\tau}_0}\sum_{n=-\infty}^{\infty}
 \mu_0^\phi\bigl(\tau_0\cdot (\Delta_0\circ T_0^n)\bigr).
\end{equation}
Recalling that $Z=4\pi(r_1+r_2)$ is the normalization of $\mu_0$, see text
after eq.\eqref{mu0}, and $Z_{\phi} =2(r_1+r_2)$ is independent of $\phi$, see
eq.\eqref{Zphi}, we have that $Z_\phi/Z=1/2\pi$. Again we see that the
fluctuations of the density $h_E$
are linear in $E$, to the leading order, and the factor of $E$ is given
by an infinite sum of correlations. The latter converges exponentially
fast according to general results (Theorem 7.31 in \cite{CM}).

Usually its central term ($n=0$) is the most significant, and it is given by

\begin{equation}  \label{scnd}
  \frac{Z_{\phi}}{Z}\frac{1}{2\bar\tau_0}\,\mu^\phi_0\bigl(\tau_0
\Delta_0\bigr)
  =\frac{\cos\phi}{4\pi\bar\tau_0}\,\mu_0^\phi\bigl(\tau_0^2\bigr).
\end{equation}

The central term explicitly involves the second moment of
$\tau_0$ restricted to $V_\phi$. Even though $\cos\phi$ has a
\emph{maximum} at $\phi=0$, it may be more than counterbalanced by a
dip that the second moment $\mu_0^\phi(\tau_0^2)$ has near $\phi=0$.
This is exactly what happens in our model shown in
Figure~\ref{figure1}.

To check numerically the above results we proceed like in the case of $n_E(\x)$
in Figure \ref{Figure_angle}. We introduce the odd part of the angular
distribution $h^o_E(\phi)=(h_E(\phi)-h_{-E}(\phi))/2$ and observe that it
satisfies the linear response equation 
\begin{equation}\label{aax}
h^o_E(\phi)=a(\phi)+o(E)
\end{equation}
with $a(\phi)$ still given by eq.(\ref{kax}). Again we expect the reminder to be
smaller. Finally, due to the symmetry of our system, we have that
$h^o_E(\phi)=(h_E(\phi)-h_E(\phi+\pi))/2$.

\begin{figure}[ht]
 \centering
   \epsfig{file=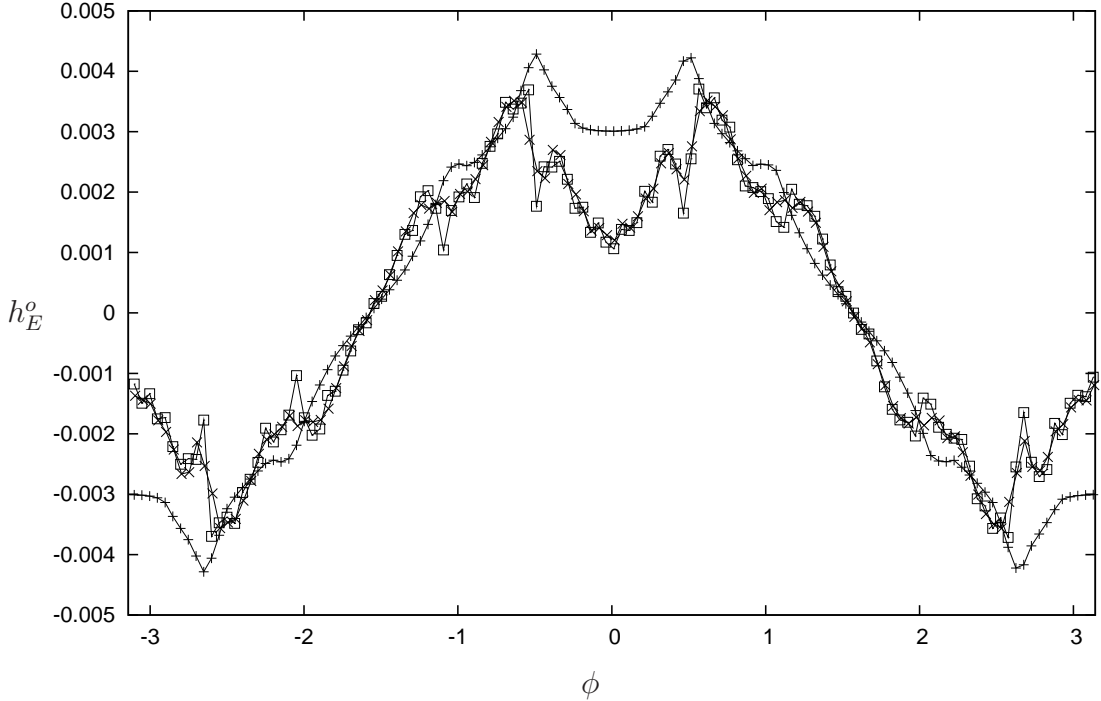, width=0.85\linewidth}
           \put(-200,-10){\makebox(0,0)[r]{$\phi$}}
      \put(-410,130){\makebox(0,0)[r]{$h^o_E$}}
  \centering
\caption{Comparison between the angular distribution for $E=0.1$ and
eq.(\ref{Kawah}); see details after eq.\eqref{scnd}.}
\label{Figure_anglec}
\end{figure}

Figure~\ref{Figure_anglec} presents the plot of eq.\eqref{aax} and the
numerically computed plot of $h^o_E(\phi)$ for $E=0.1$. The crosses
represent the numerically computed value of $h^0_E(\phi)$. The pluses come from
the central term of eq.\eqref{kax}.
Already at this level the dip is clearly visible and the agreement is
pretty good. Finally the boxes represent eq.\eqref{kax} truncated at
$|n|=20$. We have computed eq.\eqref{kax} truncating the sum up to
$|n|=100$ but no significant difference from $|n|\leq 20$ can be
observed. This is clearly consistent with a fast convergence in the sum
in eq.\eqref{kax}.

Our analysis indicates that
the dip at $\phi=0$ appears to be an artifact of the geometry of the
scatterers chosen for our deterministic model.

\section{Random Collision Models}\label{s:stocha}

In \cite{BDLR}, we introduced a simplified version of the MH model by replacing
the collisions with the fixed obstacles with a Poisson random collision
process. More precisely we assume that, in every time interval $dt$ the particle
has a probability equal to $\lambda|\p|dt$ (with $|\p|$ in this case fixed to be
1) to undergo a collision. Between collisions the particle moves according to
eq.(\ref{motion}) without $F_\mathrm{obs}$. When a collision happens we
consider two collision rules:
\begin{itemize}
\item[I] the velocity of the particle after the collision is in direction
$\phi\in[0,2\pi]$ with probability density $d\phi/2\pi$; or
\item[II] an angle $\eta\in[-\pi/2,\pi/2]$
is chosen at random with probability proportional to $\cos(\eta)d\eta$ and the
direction of the velocity is changed according to an elastic collision rules
for a particle colliding with an obstacle with outgoing velocity forming an
angle $\eta$ with the normal to the obstacle.
\end{itemize}
We call the models with the above collision rules Model I and Model II.

We can think of Model II as representing a situation in which we have $N$
scatterers with diameter $\epsilon$ randomly placed in $\mathbb{T}$ and we
consider the (Boltzmann-Grad) limit in which $N\to\infty$, $\epsilon\to 0$,
such that $N\epsilon^2\to 0$ while $N\epsilon \to\lambda^{-1}$, the mean free
path \cite{Sp}.

Let $f_\alpha(E,\x,\phi,t)$ be the probability density at time $t$ of finding
the particle at $x$ with momentum $p=(\cos\phi,\sin\phi)$. Here
$\alpha=$I,II indicates Model I or Model II respectively. This density satisfies
the equation:

\begin{eqnarray}\label{sst}
&&\partial_t f_\alpha(E,\x,\phi,t)-\p\partial_\x f_\alpha(E,\x,\phi,t) - E
\partial_\theta\left(\sin\theta f_\alpha(E,\x,\phi,t)\right)=\crcr
&&\qquad\qquad\lambda\left(\int_{-\frac{\pi}{2}}^{\frac{\pi}{2}}
p_\alpha(\eta)
f_\alpha(E,\x,\phi+2\eta+\pi,t)\,d\eta-f_\alpha(E,\x,\phi,t)\right)
\end{eqnarray}
where, $t\in\mathbb{R}^+$, $\x\in\mathbb{T}$, the unit torus, $\E=(E,0)$ is in
the horizontal
direction and $\lambda$ is the collision rate. Moreover we have
$p_\mathrm{I}(\eta)=\pi^{-1}$ for Model I and $p_\mathrm{II}(\eta)=\cos\eta/2$
for Model II.

It follows from eq.\eqref{sst}
that, when the distribution at time 0, $f_\alpha(E,x,\phi,0)$ does not depend on
$\x$,
the density $f_\alpha(E,\x,\phi,t)$ will also not depend on $\x$ for every
$t>0$. Even
if
the initial state does depend on $x$, it is easy to show \cite{BL} that as
$t\to\infty$ the system will approach a stationary density $f_\alpha(E,\phi)$
which will satisfy the equation:

\begin{equation}\label{steady}
 -\frac{E}{\lambda} \partial_\phi\left(\sin\phi
f_\alpha(E,\phi)\right)=\int_{-\frac{\pi}{2}}^{\frac{\pi}{2}}
p_\alpha(\eta)
f_\alpha(E,\phi+2\eta+\pi)\,d\eta-f_\alpha(E,\phi).
\end{equation}
From now on we will set $\lambda=1$ since the stationary $f_\alpha$ depends only
on $E/\lambda$.

\begin{figure}
  \centering
    \epsfig{file=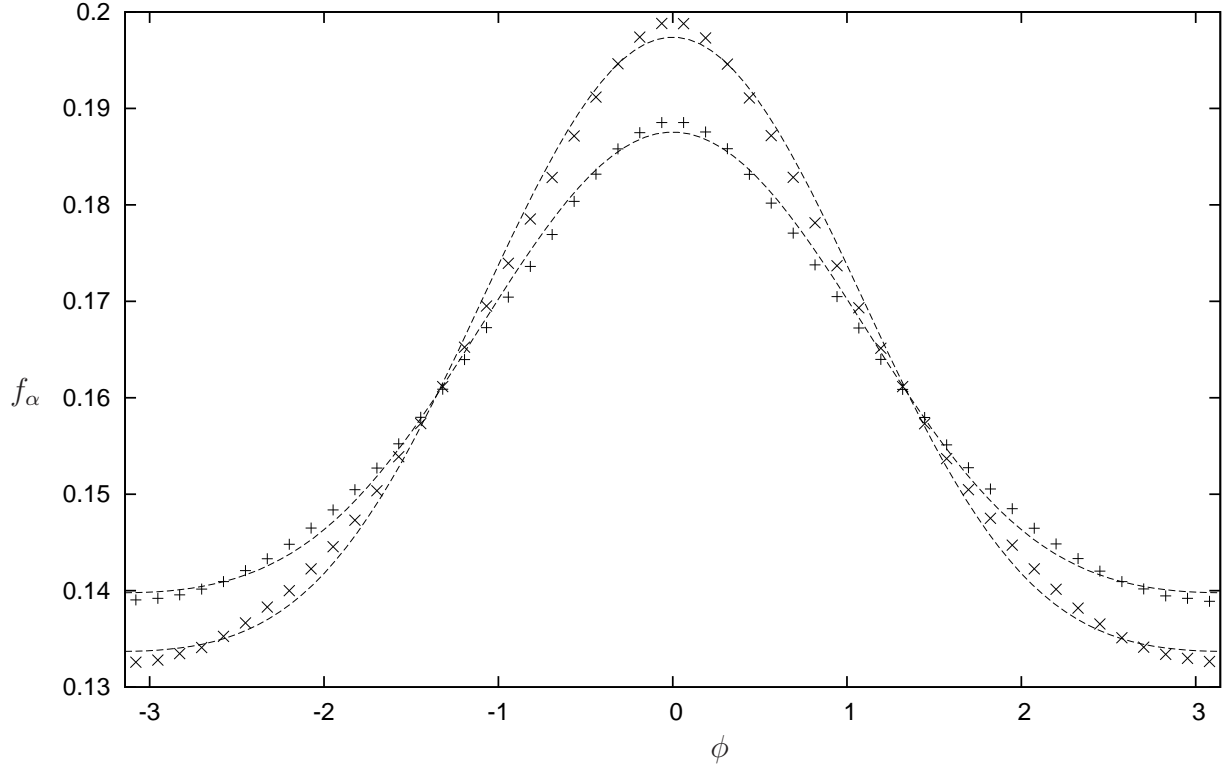, width=0.95\linewidth}
            \put(-200,-5){\makebox(0,0)[r]{$\phi$}}
      \put(-460,130){\makebox(0,0)[r]{$f_\alpha$}}
   \centering
\caption{Comparison between numerical simulations of the stochastic
process and
the power series expansion eq.(\ref{power}) for $E=0.2$. See explanation after
eq.\eqref{power}.
}
\label{Figure_stocha}
\end{figure}

We can try to solve this equation as a power series in $E$. Since $E$ is a
singular perturbation the series will not be convergent for any non zero value
of $E$. However, we expect that it will be an asymptotic series and
accurate for small $|E|$. Writing

\begin{equation}\label{powexp}
  f_\alpha(E,\phi)=\sum_{i=0}^{\infty}E^if_\alpha^{(i)}(\phi)
\end{equation}
yields a hierarchy of equations for $i=0,1,2\dots$:

\begin{equation}\label{pert}
 - \partial_\phi\left(\sin\phi
f_\alpha^{(i-1)}(\phi)\right)=\int_{-\frac{\pi}{2}}^{\frac{\pi}{2}}
p_\alpha(\eta)
f_\alpha^{(i)}(\phi+2\eta+\pi)d\eta-f_\alpha^{(i)}(\phi)
\end{equation}
with $f_\alpha^{(-1)}\equiv 0$. The equation for $i=0$ is easily solved and
gives, as the unique solution,
$f_\alpha^{(0)}\equiv 1$, since we require $\int f_\alpha(E,\phi)\,d\phi=2\pi$.
To solve the higher order equations we write

\[
f_\alpha^{(i)}(\phi)=\sum_{n=-\infty}^{\infty} \hat f_\alpha^{(i)}(n)\cos(n\phi)
\]
where we used the symmetry with respect to the direction orthogonal to the
field to eliminate the terms in $\sin(n\theta)$ and clearly
$f_\alpha^{(i)}(n)=f_\alpha^{(i)}(-n)$. In this way, for $n \neq 0$,
eq.(\ref{pert})
becomes

\begin{equation}\label{gerarchia}
 \hat f^{(i)}_\alpha(n)=\frac{n}{2}\left(1-\hat p_\alpha(n)\right)\left(\hat
f^{(i-1)}_\alpha(n-1)-\hat f^{(i-1)}_\alpha(n+1)\right)
\end{equation}
with $\hat p_\mathrm{I}(n)=0$ for Model I and $\hat p_\mathrm{II}(n)=1/4n^2$ for
Model II. Finally $f^{(0)}_\alpha(n)=\delta_{n,0}$, again due to the
normalization condition. This yields

\begin{eqnarray}\label{power}
 f_\mathrm{I}(E,\phi)&=&1+E\cos(\phi)+E^2\cos(2\phi)+\ldots\crcr
 f_\mathrm{II}(E,\phi)&=&1+\frac{3}{4}E\cos(\phi)+\frac{45}{64}
E^2\cos(2\phi)+\ldots
\end{eqnarray}
We can compare the above results with
numerical simulation of the stochastic processes generating eq.(\ref{steady}).
We set $E=0.2$ and run both processes for $10^8$ collisions. The
results are plotted in Figure \ref{Figure_stocha}. The crosses refer to Model I
while the pluses refer to Model II. Superimposed are the graph obtained from
eq.(\ref{power}). As one can see, the fit is very good. This is in agreement
with our expectation that the series in $E$ is an asymptotic one. In the case
of model I this can be rigorously justified, see eq.\eqref{devuni} below.

We note that, for both Models, the power series for  $f_\alpha(0.2,\phi)$ has a
global maximum for $\phi=0$. Since this is not true for the angular distribution
of the deterministic MH model, see section \ref{s:angular}, we will
investigate the behavior of $f(E,\psi)$ near $\phi=0$ more closely.

\subsection{Model I}

\newcommand{\fI}{f_\mathrm{I}}

Eq.(\ref{steady}) can be written as:

\begin{equation}\label{steady1}
 -E \partial_\phi\left(\sin\phi
\fI(E,\phi)\right)=
1-\fI(E,\phi)
\end{equation}
where we normalize $\fI$ as $\int \fI(E,\phi)\,d\phi=2\pi$.

Eq.(\ref{steady1}) can be solved by introducing the function

\[
 h(E,\phi)=\left(\frac{1-\cos\phi}{1+\cos\phi}\right)^{\frac{1}{2E}}
\]
which is a solution of the differential equation $\partial_\phi
h(E,\phi)=\frac{h(E,\phi)}{E\sin\phi} $, and defining
\[
 \fI(E,\phi)=\frac{h(E,\phi)}{\sin\phi}g(E,\phi)
\]
Substituting in eq.(\ref{steady1}) we obtain

\[
\partial_\phi g(E,\phi)=-h(E,\phi+\pi)
\]

Observe that $h(E,\phi)$ has a non integrable singularity at $\phi=\pi$ so
that, for $\fI(E,\phi)$ to be integrable we need $g(E,\pi)=0$. We can thus
represent the solution as:

\begin{equation}\label{soluni}
\fI(E,\phi)=\frac{h(E,\phi)}{E\sin(\phi)}\int_\phi^\pi
h(E,\eta+\pi)\,d\eta
\end{equation}
We list below some properties of $\fI(E,\phi)$ that will be useful in the
following. We have two possible situations:

\begin{itemize}
 \item[$E<1$]: In this case $\fI(E,\phi)$ is continuous in $\phi$ for every
$\phi$. Moreover it is easy to see that $\fI(E,\phi)$ is $C^\infty$
for $\phi\not =0,\pi$. For $\phi=0,\pi$, if $E<1/n$, $\fI(E,\phi)$
is $C^{n-1}$ and $\partial^n_\phi \fI(E,\phi)$ is
H\"older continuous of exponent $\alpha$ for $0<\alpha<1/E-n$.

\item[$E>1$] In this case $\fI(E,\phi)$ is still $C^\infty$
everywhere but for $\phi=0,\pi$. At $\phi=0$ we have a singularity and
$\fI(E,\phi)\simeq \phi^{1/E-1}$. More precisely the function
$\phi^{1-1/E'}\fI(E,\phi)$ is H\"older continuous of exponent
$0<\alpha<1/E-1/E'$, for every $E'>E$.
\end{itemize}

Starting form eq.(\ref{soluni}) and integrating by part we obtain:

\begin{eqnarray}\label{devuni}
\fI(E,\phi)&=&1-\frac{h(E,\phi)}{\sin(\phi)}\int_\phi^\pi\cos\eta\,
h(E,\eta+\pi)\,d\eta=\crcr
&=&1+E\cos\phi+E\frac{h(E,\phi)}{\sin(\phi)}\int_\phi^\pi\cos2\eta\,
h(E,\eta+\pi)\,d\eta=\crcr
&=&1+E\cos\phi+E^2\cos(2\phi)+E^2\frac{h(E,\phi)}{\sin(\phi)}
\int_\phi^\pi\partial_\phi[\cos2\eta\sin\eta]h(E,\eta+\pi)\,d\eta=\crcr
&=&\sum_{i=0}^{N}E^n\fI^{(i)}(\phi)+E^N R_N(E,\phi)
\end{eqnarray}
The above expansion coincides with the one obtained in
eqs.(\ref{powexp}-\ref{gerarchia}).
It is not difficult to see that $|R_N(E,\phi)|\leq KC^N N!$. Since it is
clear from eq.\eqref{soluni} that $f_I(E,\phi)$ is not analytic in $E$ for
small $E$, this inequality means
that, as we discussed previously, the perturbative series for $\fI(E,\phi)$ is
at least asymptotic. Notwithstanding this,
eq(\ref{devuni}) and the regularity properties of $\fI(E,\phi)$ tell us
that, for $E$ small, $\fI(E,\phi)$ has a unique maximum at $\phi=0$ and a
unique minimum at $\phi=\pi$.

\subsection{Model II}

\newcommand{\fII}{f_\mathrm{II}}

We can use the solution of Model I to get more analytical information on Model
II. Proceeding as in eq.(\ref{steady1}) we write the solution of
eq.(\ref{steady}) as
\[
 \fII(E,\phi)=\frac{h(E,\phi)}{E\sin\phi}g(E,\phi)
\]
and obtain the representation for $g(E,\phi)$:

\begin{equation}
 \partial_\phi
g(E,\phi)=-\frac{1}{4E}h(E,\phi+\pi)
\int_{-\pi}^{\pi}
\left|\cos\left(\frac { \omega-\phi }
{2}\right)\right|\frac{h(E,\omega+\pi)}{\sin(\omega+\pi)}g(E,\omega+\pi)\,
d\omega
\end{equation}
from which, reasoning as in Model I, we get

\begin{equation}\label{Perron}
 \fII(E,\phi)=\frac{1}{4E\sin(\phi)}h(E,\phi)\int_\phi^\pi
h(E,\eta+\pi)\int_{-\pi}^{\pi}\left|\cos\left(\frac{
\omega-\eta } {2}\right)\right|f(E,\omega+\pi)\,d\omega\, d\eta
\end{equation}
for $0<\phi<\pi$. We can then set $\fII(E,-\phi)=\fII(E,\phi)$.
Observe that the above equation can be written has as

\begin{equation}\label{PF}
 \fII(E,\phi)=\int_{-\pi}^\pi Q(\phi,\omega)\fII(E,\omega)\,d\omega
\end{equation}
where
\[
 Q(\phi,\omega)=\frac{1}{4E\sin(\phi)}h(E,\phi)\int_\phi^\pi
h(E,\eta+\pi)\left|\sin\left(\frac{\omega-\eta } {2}\right)\right|\,d\eta
\]
for $0<\phi<\pi$ and $Q(-\phi,\omega)=Q(\phi,\omega)$.
It is easy to see that $Q(\phi,\omega)>0$ for every $\phi,\omega$. Moreover we
have
\begin{eqnarray}
 \int_0^\pi Q(\phi,\omega)\,d\phi&=&\lim_{\epsilon\to
0}\int_\epsilon^{\pi-\epsilon} Q(\phi,\omega)\,d\phi=\crcr
&=&\frac{1}{4}\int_0^\pi h(E,\phi)h(E,\phi+\pi)
\left|\sin\left(\frac{\omega-\phi } {2}\right)\right|\,d\phi+\crcr
&-&\frac{1}{4}\lim_{\epsilon\to 0}h(E,\epsilon)\int_\epsilon^\pi
h(E,\eta+\pi)\left|\sin\left(\frac{\omega-\eta } {2}\right)\right|\,d\eta\crcr
&+&\frac{1}{4}\lim_{\epsilon\to 0}h(E,\pi-\epsilon)\int_{\pi-\epsilon}^\pi
h(E,\eta+\pi)\left|\sin\left(\frac{\omega-\eta } {2}\right)\right|\,d\eta=\crcr
&=&\frac{1}{4}\int_0^\pi
\left|\sin\left(\frac{\omega-\phi } {2}\right)\right|\,d\phi
\end{eqnarray}
where we have used that $h(E,\phi)=h(E,\phi+\pi)^{-1}$ and that, for $\epsilon$
small, $h(E,\epsilon)\simeq \epsilon^{1/E}$ while  $h(E,\epsilon+\pi)\simeq
\epsilon^{-1/E}$. Proceeding in the same way for $-\pi<\phi<0$, we get
\[
 \int_{-\pi}^\pi Q(\phi,\omega)\,d\phi=\frac{1}{4}\int_{-\pi}^\pi
\left|\sin\left(\frac{\omega-\phi } {2}\right)\right|\,d\phi= 1
\]
for every $\omega$. Finally, in the same way we
got the regularity properties of $\fI(E,\phi)$, we can see that, if $E<1$
then $Q\fII$ is a H\"older continuous function with H\"older norm bounded by the
$L^\infty$ norm of $\fII$. This immediately implies, by the Ascoli-Arzel\'a
theorem, that $Q$ is a compact linear operator on $C^0$. In this situation
we can apply the Krein-Rutman theorem, see \cite{KR,Di}, and obtain that there
is a unique function $\fII(E,\phi)$ that satisfies eq.(\ref{PF}). Moreover
$\fII(E,\phi)>0$ for every $\phi$. A similar argument tells us that, for $E>1$,
there is a unique solution of eq.(\ref{PF}) and it can be written as
\[
\fII(E,\phi)=|\sin(\phi)|^{1-\frac{1}{E}}l(E,\phi)
\]
with $l(E,\phi)$ continuous in $\phi$ and strictly positive.

Observe that, for any integrable function $f(\phi)$, we have

\begin{eqnarray}
\partial_\phi \int_{-\pi}^{\pi}\left|\cos\left(\frac{\omega-\phi }
{2}\right)\right|f(\omega+\pi)d\omega&=&
\frac{1}{2}\int_{-\pi}^{\pi}
{\rm sgn}(\omega-\phi)\left|\sin\left(\frac{\omega-\phi}
{2}\right)\right|f(E,\omega+\pi)d\omega\crcr
\partial_\phi^2 \int_{-\pi}^{\pi}\left|\cos\left(\frac{
\omega-\phi}{2}\right)\right|f(E,\omega+\pi)d\omega&=&
-\frac{1}{4}\int_{-\pi}^{\pi}\left|\cos\left(\frac{
\omega-\phi } {2}\right)\right|f(E,\omega+\pi)d\omega+f(\phi)\nonumber
\end{eqnarray}
Thus the above integral is always at least $C^1$, while if $f(\phi)$ is
 $C^n$ it is $C^{n+2}$. This implies that $\fII(E,\phi)$ has the same regularity
properties as a function of $\phi$ as $\fI(E,\phi)$. In particular if $E\leq
1/3$, $\fII(E,\phi)$ is $C^2$ and we can try to compute
$\fII''(E,0)$ explicitly. Observe that eq.(\ref{steady}) tells us that

\begin{equation}
E\sin\phi \fII'(E,\phi)+E\cos\phi \fII(E,\phi)=
-\frac{1}{4}\int_{-\pi}^{\pi}\left|\cos\left(\frac{
\omega-\phi}{2}\right)\right|\fII(E,\omega+\pi)d\omega+\fII(E,\phi)
\end{equation}
Evaluating at $\phi=0$ we get
\begin{equation}\label{at0}
 \fII(E,0)=\frac{1}{4(1-E)}\int_{-\pi}^{\pi}\cos\left(\frac{
\omega}{2}\right)\fII(E,\omega+\pi)d\omega
\end{equation}
As expected, this equation loose meaning when $E\geq 1$ since $\fII(E,\phi)$ is
no more continuous at $\phi=0$. We can now differentiate both side of
eq.(\ref{steady}) and obtain, after evaluating in $\phi=0$,
\[
 (1-2E)\fII'(E,\phi)=\frac{1}{8}\int_{-\pi}^{\pi}\sin\left(\frac{
\omega}{2}\right)\fII(E,\omega+\pi)d\omega=0
\]
for symmetry reasons. Again this equation make sense only if $E<1/2$. Finally,
differentiating once more, we get

\[
3E\fII''(E,0)-E\fII(E,0)=
\frac{1}{16} \int_{-\pi}^{\pi}\cos\left(\frac{
\omega}{2}\right)\fII(E,\omega)d\omega-\frac{1}{4}\fII(E,0)+\fII''(E,0)
\]
Using eq.(\ref{at0}) we get
\begin{equation}\label{at02}
 \fII''(E,0)=-\frac{3}{4}\frac{E}{1-3E}\fII(E,0)
\end{equation}
that is clearly negative for $E<1/3$ so that we have that $f(E,\phi)$ has a
local maximum at $\phi=0$. Observe that expanding this formula to third order
in $E$ we get a result in agreement with the expansion in eq.(\ref{power})

\section*{Acknowledgment}
The authors thank Eric Carlen for many insightful comments and
discussions. The work of FB was supported in part by NSF grant 0604518. The work
of NC was supported in part by NSF grant DMS-0969187. The work of JLL was
supported in part by NSF grant DMR08021220 and by AFOSR grant AF-FA9550-07. The
authors are also grateful to the Alabama supercomputer administration for
computational resources.

\renewcommand{\appendixname}{Appendix}
\begin{appendices}
\appendixpage
\renewcommand{\appendixname}{Appendix}

\section{Regularity of Projections of SRB measures}
\label{a:regu}

SRB measures are characterized by absolutely continuous conditional
distributions on unstable manifolds, but generally they are singular.
Singularity of a measure $\mu$ means that there is a subset $\M' \subset
\M$ in the phase space $\M$ such that $\mu(\M')=1$ but the Lebesgue measure
of $\M'$ is zero. In that case $\mu$ does not have a density on $\M$.

However in physics one rarely observes measures on the entire phase
space; it is more common to observe distributions of some selected
variables (e.g., positions or velocities of selected particles). The
distributions of those variables are obtained by projection of the
relevant measure onto the corresponding variables. And the resulting
distribution is often absolutely continuous, with a continuous
density, despite the singularity of the measure in the whole of phase space.

Similar smoothness results hold if, instead of projecting the measure
onto certain variable(s), we integrate some smooth functions with
respect to all the other variables. For example, in eq.\eqref{nunu} we
integrate $\cos \phi$ and $\sin \phi$ with respect to $\phi$ and get an
absolutely continuous distribution with respect to $\x$ with a
continuous density.

We sketch a proof here that in our Moran-Hoover model the corresponding
projections have continuous bounded densities. Let $\mu_E$ denote the SRB
measure on the collision space with coordinates $(\theta,\psi)$ for a given
value of $E$. Since we will consider only a given value of $E$, we will
suppress the dependence on $E$ in what follows. All estimates are uniform
in $E$. Consider
a projection of $\mu$ onto a line transversal (not parallel or
perpendicular) to stable and unstable manifolds and singularity
manifolds. For simplicity, let $\mu$ be projected onto the $\theta$ axis.

The density of the projection can be computed as
$$
   \rho(\theta) = \lim_{\delta \to 0} \delta^{-1} \mu(R_{\theta,\delta})
$$
where $R_{\theta,\delta} = \{(\theta',\psi)\colon
\theta<\theta'<\theta+\delta\}$ is a rectangle in the collision space of
size $\delta$ in the $\theta$ direction. It is known that the SRB measure
$\mu$ satisfies
\begin{equation} \label{Cd}
       \mu(r(\theta,\psi)<\delta) < C\delta\qquad \forall \delta>0
\end{equation}
for some constant $C>0$ which is uniform for all small fields $E$, see
\cite{Ch}. Here $r(\theta,\psi)$ denotes
the distance from $(\theta,\psi)\in \M$ to the nearer endpoint of the unstable
manifold
passing through $(\theta,\psi)$ (that manifold is a smooth curve which is
divided
by $(\theta,\psi)$ into two segments; so $r(\theta,\psi)$ denotes the length of
the shorter
one).

Now for each unstable manifold $W$, the intersection $W\cap
R_{\theta,\delta}$ is a segment of $W$ that has length bounded above by $C'
\delta$ for
some constant $C'>0$. Hence its relative measure (within $W$) is of the
same order as the measure of the segment of length $\delta$ at an
endpoint of $W$. We recall that the conditional densities of SRB
measures are H\"older continuous, and their fluctuations are uniformly
bounded \cite{CM,Ch,Z}. Now eq.\eqref{Cd} implies
\[
       m(R_{\theta,\delta}) < C''\delta
\]
for some constant $C''>0$, hence $\rho(\theta)$ is uniformly bounded.

Next we prove that $\rho(\theta)$ is continuous. For $\theta_1\approx
\theta$ we have
\[
   \rho(\theta_1)-\rho(\theta) =
   \lim_{\delta \to 0}
   \frac{\mu(R_{\theta_1,\delta})-\mu(R_{\theta,\delta})}{\delta}
\]
Now there are unstable manifolds that cross both rectangles
$R_{\theta_1,\delta}$ and $R_{\theta,\delta}$ and those which cross
only one of them; accordingly we have
\[
   \mu(R_{\theta_1,\delta})-\mu(R_{\theta,\delta}) = \Delta_1+\Delta_2,
\]
where $\Delta_1$ accounts for the former, and $\Delta_2$ for the
latter. The conditional density of $\mu$ on each unstable manifold is
H\"older continuous, and unstable manifolds have uniformly bounded
curvature \cite{Ch,Z}, hence once can easily see that
\begin{equation}  \label{Delta1}
    |\Delta_1| \leq C\delta|\theta_1-\theta|^\gamma
\end{equation}
for some constants $C>0$ and $\gamma>0$ (in fact, $\gamma=1/3$ for our
model; cf.\ \cite[Corollary 5.30]{CM}). It remains to show that
\begin{equation} \label{ooo}
   \lim_{\theta_1\to\theta} \limsup_{\delta \to 0}
     |\Delta_2|/\delta=0,
\end{equation}
i.e., the contribution from unstable manifolds crossing just one
rectangle is negligible.

To estimate $\Delta_2$, denote by $F(y) = \mu(\cup W\colon |W|<y)$ be the
measure of all the unstable manifolds of length $<y$. Then eq.\eqref{Cd}
can be written as
\[
  F(2\delta) + \int_{2\delta}^{\infty} \frac{2\delta}{y}\, dF(y) \leq C\delta.
\]
Dividing by $\delta$ and taking the limit $\delta\to 0$ gives
\begin{equation} \label{i1}
  2 \int_0^{\infty} \frac{dF(y)}{y} < C
\end{equation}
(see also \cite[Exercise 7.15]{CM}). Now let $L_{\theta} =
\{(\theta,\psi)\colon \psi\in[-\pi/2,\pi/2]\}$ denote the line in $\M$ with
the fixed $\theta$ coordinate. Denote
\[
  F_{\theta,\theta_1}(y) = \mu(\cup W\colon |W|<y,\ W\text{ terminates between }
  L_{\theta}\text{ and }L_{\theta_1}).
\]
Then we have
\begin{equation} \label{i2}
  \limsup_{\delta \to 0}
  \frac{|\Delta_2|}{\delta} \leq \int_0^{\infty}
\frac{dF_{\theta,\theta_1}(y)}{y}
\end{equation}
Now, as $\theta_1$ and $\theta$ get closer together,
$F_{\theta,\theta_1}(y)$ monotonically decreases for each fixed $y>0$.
Moreover, we have
\begin{equation} \label{i3}
   \lim_{\theta_1\to \theta} F_{\theta,\theta_1}(y) = 0\qquad \forall y>0
\end{equation}
because the union of unstable manifolds terminating exactly on the line
$L_{\theta}$ has $\mu$-measure zero. To see this observe that unstable manifolds
terminate on singularity lines of the past iterations of the collision map
$T$, i.e., on singularity lines of $T^{-n}$, $n>0$. These lines intersect
the line $L_{\theta}$ at countably many points thus there are at most
countably many unstable manifolds terminating on $L_{\theta}$. Finally
each individual unstable manifold has $\mu$-measure zero, as is
guaranteed by the Poincar\'e recurrence theorem. Now combining
eq.\eqref{i1}--eq.\eqref{i3} proves eq.\eqref{ooo}.

In smooth hyperbolic systems without singularities all unstable
manifolds are long enough so that $\Delta_2=0$. Then the density $\rho$ is
not only continuous, but H\"older continuous, according to
eq.\eqref{Delta1}. We believe that in our MH model, too, the main
contribution to the structure of $\rho$ comes from $\Delta_1$, so that
$\rho$ is also H\"older continuous, but our estimate eq.\eqref{ooo} on
$\Delta_2$ is too poor to prove that.

The above argument applies to projections of SRB measures onto some
coordinates (transversal to stable and unstable directions). If,
instead of projections, we integrate smooth functions like in
eq.\eqref{nn}, then those functions can be incorporated into conditional
densities of the SRB measure on unstable manifolds, and the argument
will work for that situation, too.

\section{Derivation of Green-Kubo formulas}

The derivation of eqs.(\ref{dx}-\ref{nx}) and (\ref{ax}) is based on a
Kawasaki-type formula
used in linear response theory, see \cite{CELS}. For a small external
field $E$ and the corresponding SRB measure $\mu_E$, we can integrate
any bounded piecewise H\"older continuous function $f_E$ on $\M$ as
follows:
\begin{equation} \label{Kawa0a}
  \mu_E(f_E) = \mu_0(f_E) + \sum_{n=1}^{\infty}
  \mu_0\bigl((f_E\circ T_E^n)(1-e^{-E\Delta_E})\bigr),
\end{equation}
where $\Delta_E$ denotes the displacement of the particle in the
direction of the field (i.e., in the positive $x$ direction) during its
free flight to the next collision. More precisely, for $(\theta,\psi)
\in \M$ we set
\begin{equation}\label{DD}
 \Delta_E(\theta,\psi)=\int_0^{\tau_E(\theta,\psi)}\cos(\Phi_t^E(\theta,\psi))\,
 dt
\end{equation}
where we use this definition to avoid the ambiguity on the difference between
two points on a torus (periodic boundary conditions).
The above Kawasaki-type formula eq.\eqref{Kawa0a} is derived in
\cite[Eq.~(16)]{CELS}.

The function $f_E$ may depend on the field $E$, but it must have a
limit $f_0 = \lim f_E$ as $E\to 0$. Since $T_0 = \lim_{E\to0} T_E$
and $\tau_0 = \lim_{E\to0}  \tau_E$, as well as $\Delta_0 = \lim_{E\to0}
\Delta_E=\tau\cos\phi$, a first order Taylor expansion of the infinite sum
in eq.(\ref{Kawa0a}) gives
\begin{equation} \label{Kawa1a}
  \mu_E(f_E) = \mu_0(f_E) + E\sum_{n=1}^{\infty}
  \mu_0\bigl((f_0\circ T_0^n)\Delta_0\bigr) + o(E),
\end{equation}
see \cite[Eq.~(17)]{CELS}. The infinite sum in the above equation
converges because $\mu_0\bigl((f_0\circ
T_0^n)\Delta_0\bigr)\to \mu_0(f_0)\mu_0(\Delta_0)=0$ exponentially
fast as $n\to\infty$ for $f_0$ H\"older continuous (decay of correlations)
and $\mu_0(\Delta_0)=0$.
Note that this sum is independent of $E$, hence the second
term is linear in $E$. The first term $\mu_0(f_E)$ will be handled
separately for each $f_E$.

\subsection{Derivation of Eqs.(\ref{dx}) and (\ref{nx})}

We first derive eq.(\ref{dx}) for the local density $n_E(\x)$, which can be
represented, according to eq.\eqref{deltaA}, by
\begin{equation} \label{d2nx}
   \delta^2n_E(\x) =
   \mu_E (J_A^E) + o(\delta^2)
\end{equation}
where $A = \{\x'\colon \|\x'-\x\|_{\infty} \leq \delta/2\}$ denotes the
square in $Q$ with side $\delta$ centered on $\x$. We fix a small
$\delta>0$ and will estimate the integral in eq.\eqref{d2nx}.

It is convenient to
extend the space $\M$ by adding the sides of the square $A$ to $\partial
Q$. In other words, every time a trajectory crosses the boundary of $A$ and
enters or exits $A$
we register a `virtual collision' (the trajectory does not actually
change its direction, so the collision has no effect on the trajectory of the
particle, {\sl i.e.} $\partial
A$ plays the role of `transparent walls'). Adding transparent walls with
virtual collisions is a useful trick in the study of billiards.

By adding this transparent wall the phase space of the system is thus
extended from $\M=[0,4\pi]\times [-\pi/2,\pi/2]$ to
$\M_A=[0,4\pi+4\delta]\times [-\pi/2,\pi/2]$ where
$\theta\in[4\pi,4\pi+4\delta]$ parametrize $\partial A$. Consistently we
must replace the map $T_E$ with the new map $T_{E,A}$ constructed as in
section \ref{intro}, the SRB measure $\mu_E$ with $\mu_{E,A}$ defined as in
eq.(\ref{wlim}) and the function $J_A^E$ with $J_A^{E,A}$ defined as in
eq.(\ref{JA}). The notation $J_A^{E,A}$ helps keeping track of the fact
the $A$ appears both in the indicator function appearing in eq.(\ref{JA})
and in the phase space $\M_A$ on which $J_A^{E,A}$ is defined. From this
follows that $J_A^{E,A}\neq 0$ if and only if $\theta$ represents a collision
taking place on $\partial A$ and the outgoing velocity points inside $A$.
Indeed, $I_A({\bf X}_t(\theta,\psi))\not = 0$ if and only if ${\bf
X}_t(\theta,\psi)\in A$. But in this case the last collision of the trajectory
was with $\partial A$ and the velocity was pointing inside $A$.

Clearly eq.\eqref{Kawa1a} remains true if we replace $\mu_E$ with $\mu_{E,A}$.
We apply it to $f_E = J_A^{E,A}$ and get
\begin{equation} \label{Kawa2a}
  \delta^2n_E(\x) = \mu_{0,A}( J_A^{0,A}) + \mu_{0,A}(\chi_A^E) +
E\sum_{n=1}^{\infty}
  \mu_{0,A}\bigl(( J_A^{0,A}\circ T_{0,A}^n)\Delta_{0,A}\bigr) + o(E)
\end{equation}
where $\Delta_{0,A}$ is defined as in eq.(\ref{DD}) and $\chi_A^E = J_A^{E,A} -
J_A^{0,A}$. By direct calculations we get
$$
  \mu_{0,A}( J_A^{0,A}) = \frac{\delta^2}{{\rm Area}(Q)} = \delta^2 n_0(\x).
$$
Observe that, for every $E$, $J_A^{E,A}=O(\delta)$ while $\mu_{0,A}({\rm
supp}( J_A^{E,A}))=O(\delta)$. Moreover, from eq.(\ref{motion}) we have
that $|{\bf X}^E_t(\theta,\psi)-{\bf X}^0_t(\theta,\psi)|=O(Et^2)$ so that
$ J_A^{E,A} - J_A^{0,A} = O(\delta^2 E)$. Thus even
though the term
$\mu_{0,A}(\chi_A^E)=O(E\delta^3)$ is linear in $E$, its contribution vanishes
in the
limit $\delta\to 0$ and we will ignore it. We finally arrive at
\begin{equation} \label{nE}
  n_E(\x) = n_0(\x) +
  \delta^{-2}E\sum_{n=1}^{\infty}
  \mu_{0,A}\bigl((J_A^{0,A}\circ T_{0,A}^n)\Delta_0\bigr) + o(E).
\end{equation}
Observe that a trajectory originating from a point $\x$ gives a non zero
contribution to the $n$-th term in the sum only if its $n$-th collision is with
$\partial A$ and the trajectory enters $A$.
Since $E$ and $\delta$ are small, this implies that the $n-1$-th collision was
with a non-virtual obstacle. We can thus rewrite eq.(\ref{nE}) as

\begin{equation} \label{nEE}
  d(\x) = \delta^{-2}\left(\mu_{0}\bigl(J_A^0\Delta_{0,A}\bigr)+
  \sum_{n=1}^{\infty}\mu_{0}\bigl(( J_A^0\circ T_{0}^n)\Delta_0\bigr)\right)
\end{equation}
where we have neglected the trajectories that collide more than once with $A$,
since they contribute $O(\delta^3)$ to the integral, and the difference between
$\bar\tau_E$ and $\bar\tau_{E,A}$, appearing in $ J^{E,A}_A$, since it is
$O(E)$ and thus does not contribute at first order.

We still have to discuss the limit $\delta\to 0$ in eq.(\ref{nEE}). This limit
is non trivial since, although the correlations appearing in the infinite sum
decay exponentially for every $A$, we need to show that such a decay is uniform
in $\delta$. To show this we can take the limit $\delta\to 0$ term by term in
the sum. Taking this into account we obtain eq.(\ref{dx}).

We can now derive eq.(\ref{nx}) for the local average velocity $\n_E(\x)$.
Similarly to eq.\eqref{d2nx} we have
\begin{equation} \label{d2nnx}
   \delta^2\n_E(\x) =
   \tfrac{1}{n_E(\x)}\,\mu_E (\bH_A^E) + o(\delta^2),
\end{equation}
where
\begin{equation}\label{HH}
 \bH_A^E(\theta,\psi)=\frac{1}{\bar{\tau}_E}
\int_0^{\tau_E(\theta,\psi)}
\bigl(\cos(\Phi^E_t(\theta ,
\psi)),\sin(\Phi^E_t(\theta ,
\psi))\bigr)I_{A}\left(X^E_t(\theta ,
\psi)\right)dt
\end{equation}
and $A$ again denotes the square in $Q$ with side $\delta$ centered on
$\x$. In this case we will not need to introduce virtual collisions with
$A$ like we did for eq.\eqref{dx}. Applying eq.\eqref{Kawa0a} we obtain
\begin{equation} \label{Kawa3a}
  \delta^2n_E(\x)\n_E(\x) = \mu_0(\bH_A^E) + E\sum_{n=1}^{\infty}
  \mu_0\bigl((\bH_A^0\circ T_0^n)\Delta_0\bigr) + o(E)
\end{equation}
To eliminate the first term $\mu_0(\bH_A^E)$ we apply an
antisymmetrization. Due to the invariance of $\mu_E$ we have
$\mu_E(\bH_A^E) = \mu_E(\bH_A^E\circ T_E^{-1})$, hence, by applying
eq.\eqref{Kawa0a} to $\bH_A^E\circ T_E^{-1}$, we get
\begin{equation} \label{Kawa4a}
  \delta^2n_E(\x)\n_E(\x) = \mu_0(\bH_A^E\circ T_E^{-1}) + E\sum_{n=0}^{\infty}
  \mu_0\bigl((\bH_A^0\circ T_0^n)\Delta_0\bigr) + o(E)
\end{equation}
Next, let $\J\colon \M\to \M$ denote an involution defined by
$\J(\theta,\psi) = C(\theta,-\psi)$. Due to the time reversibility of the
perturbed dynamics we have $\J \circ T_E = T_E^{-1} \circ \J$, and
therefore $\bH_A^E = - \bH_A^E \circ \J \circ T_E$. Also note that $\mu_0$
is invariant under both $T_0$ and $\J$. Thus if we add eq.\eqref{Kawa3a}
and eq.\eqref{Kawa4a}, the first terms cancel out.

Moreover, the time reversibility of the billiard dynamics implies
$\Delta_0 = - \Delta_0 \circ \J \circ T_0$ and $\bH_A^0 \circ T_0^{n} =
-\bH_A^0 \circ T_0^{-n}\circ \J\circ T_0$ for all $n$. Therefore,
$$
   \bigl(\bH_A^0\circ T_0^{n}\bigr)\cdot \Delta_0 =
    \bigl[\bigl(\bH_A^0\circ T_0^{-n}\bigr)\Delta_0\bigr] \circ \J\circ T_0.
$$
Thus adding eq.\eqref{Kawa1} and eq.\eqref{Kawa2} together gives
$$
  \delta^2n_E(\x)\n_E(\x) = \tfrac 12 \, E\sum_{n=-\infty}^{\infty}
  \mu_0\bigl((\bH_A^0\circ T_0^n)\Delta_0\bigr) + o(E)
$$
Taking into account eq.\eqref{nE} we get
\begin{equation} \label{Kawa5a}
  \kk_E(\x) = \tfrac{1}{2\delta^2n_0(\x)} \, \sum_{n=-\infty}^{\infty}
  \mu_0\bigl((\bH_A^0\circ T_0^n)\Delta_0\bigr)
\end{equation}
where, due to the time reversibility of the dynamics, the term for $n$ and $-n$
in the sum are equal. To obtain eq.(\ref{nx}) we have used that
$\lim_{\delta\to0 }\delta^{-2}\mu_0\bigl((\bH_A^0\circ
T_0^n)\Delta_0\bigr)=c\bigl(\rho^c_n(\Delta_0),\rho^s_n(\Delta_0)\bigr)$ with
$c=n_0(\x)$.

\subsection{Derivation of Eq.(\ref{ax})} \label{derax}

To derive eq.\eqref{ax}, we apply eq.(\ref{Kawa1a}) to
$f_E=J^E_{A(\phi,\delta)}$, which was defined in eq.\eqref{JA} and where
$A(\phi,\delta)$ is the set $\phi-\delta/2 \leq \theta+\psi \leq
\phi+\delta/2$, {\sl i.e.} the set of velocity vectors that
make an angle $\phi' \in [\phi-\delta/2,\phi+\delta/2]$. Note that when
$E=0$, the trajectory is a straight line, so $f_0 =\tau_0
I_{A(\phi,\delta)}$.

As in eq.(\ref{d2nx}) we can write $\delta
h_E(\phi)=\mu_E(J^E_{A(\phi,\delta)})+o(\delta)$. We denote
$\Delta_{\phi,\delta} = \Delta_0\cdot I_{A(\phi,\delta)}$ and recall
that $\Delta_0=\tau_0\cos\phi$, so that $J^0_{A(\phi,\delta)}\cos\phi =
\Delta_{\phi,\delta}/\bar\tau_0$. Thus eq.(\ref{Kawa1a}) becomes

\begin{equation} \label{Kawa1b}
  \delta \cos\phi \, h_E(\phi) =
\mu_0(J^E_{A(\phi,\delta)})\cos\phi
    + \frac{E}{\bar\tau_0}\sum_{n=1}^{\infty}
  \mu_0\bigl((\Delta_{\phi,\delta}\circ T_0^n)\Delta_0\bigr)
\end{equation}
and, again due to the invariance of $\mu_E$, we get
\begin{equation} \label{Kawa1}
  \delta \cos\phi \, h_E(\phi) = \mu_0(J^E_{A(\phi,\delta)}\circ
T_E^{-1})\cos\phi
    + \frac{E}{\bar\tau_0}\sum_{n=0}^{\infty}
  \mu_0\bigl((\Delta_{\phi,\delta}\circ T_0^n)\Delta_0\bigr)
\end{equation}
and (denoting, for brevity, $\phi^- = \phi+\pi$)
\begin{equation} \label{Kawa2}
  \delta \cos\phi^- \, h_E(\phi^-) =
\mu_0(J^E_{A(\phi^-,\delta)})\cos\phi^-
    + \frac{E}{\bar\tau_0}\sum_{n=1}^{\infty}
  \mu_0\bigl((\Delta_{\phi^-,\delta}\circ T_0^n)\Delta_0\bigr),
\end{equation}
In all the above formulas we have suppressed the $o(E\delta)$ terms.
Time reversibility implies $J^E_{A(\phi,\delta)}\circ T_E^{-1} =
J^E_{A(\phi^-,\delta)} \circ \J$, where $\J$ denotes the time reversal
involution. Thus, if we add eq.\eqref{Kawa1} and
eq.\eqref{Kawa2}, their first terms cancel out. Similarly, we get
$\Delta_0 = - \Delta_0 \circ \J \circ T_0$ and $\Delta_{\phi^-,\delta}
\circ T_0^{n} = -\Delta_{\phi,\delta}\circ T_0^{-n} \circ \J\circ T_0$
for all $n$. Therefore,
$$
   \bigl(\Delta_{\phi^-,\delta}\circ T_0^{n}\bigr)\cdot \Delta_0 =
    \bigl[\bigl(\Delta_{\phi,\delta}\circ T_0^{-n}\bigr)\Delta_0\bigr] \circ
\J\circ T_0.
$$
Thus adding eq.\eqref{Kawa1} and eq.\eqref{Kawa2} gives

\begin{equation}\label{Kawah}
   h_E(\phi)-h_E(\phi+\pi)
   = \frac{E}{\delta\bar \tau_0\cos\phi} \sum_{n=-\infty}^{\infty}
  \mu_0\bigl((\Delta_{\phi,\delta}\circ T_0^n)\Delta_0\bigr)+o(E).
\end{equation}
This is an infinite sum of correlations which decay exponentially fast
\cite{CM}.

In the case of the billiard shown in Figure~\ref{figure1} we have that,
due to the symmetry of the system, $h_{E}(\phi+\pi)=h_{-E}(\phi)$ so
that we get
\begin{align*}
 a(\phi)&=\frac{1}{2\delta\bar \tau_0\cos\phi} \sum_{n=-\infty}^{\infty}
  \mu_0\bigl((\Delta_{\phi,\delta}\circ T_0^n)\Delta_0\bigr))\\
  &=\frac{1}{2\delta\bar \tau_0\cos\phi} \sum_{n=-\infty}^{\infty}
  \mu_0\bigl(\Delta_{\phi,\delta}\cdot(\Delta_0\circ T_0^n)\bigr)
\end{align*}
where we used the invariance of $\mu_0$ under $T_0$. Finally using the
relation $\Delta_{\phi,\delta} = \tau_0 I_{A(\phi,\delta)} \cos\phi$ and taking
the
limit $\delta\to 0$ gives eq.\eqref{kax}. 

\section{Isotropy of the Collision Time}\label{a:tau}

As a motivation for our definition of $\mu_0^\phi(\tau_0)$, we
define a \emph{directional} mean free time as follows. Given $\phi \in
[0, 2\pi]$ and $\delta>0$, let $I=I_{A(\phi,\delta)}$ be as in
Section~\ref{derax}. Due to ergodicity, we have
\begin{equation} \label{III}
   \lim_{n\to\infty} \frac {\sum_{i=0}^{n-1}
   \tau_0(T_0^i(\theta,\psi))I(T_0^i(\theta,\psi))}{\sum_{i=0}^{n-1}
   I(T_0^i(\theta,\psi))} = \frac{\mu_0(\tau_0 I)}{\mu_0(I)}
\end{equation}
 for almost every $(\theta,\psi)\in \M$, and we call the limit (if
it exists)
\[
  \bar{\tau}_\phi = \lim_{\delta\to 0} \frac{\mu_0(\tau_0 I)}{\mu_0(I)}
\]
the \emph{directional mean free time} (corresponding to the angle
$\phi$). Now arguing as in Section~\ref{derax} we get
\begin{equation} \label{area}
  \lim_{\delta\to 0} \delta^{-1}\mu_0(\tau_0 I) =
  \int\tau(\phi-\theta,\theta)\cos(\phi-\theta)\chi(\theta)\, d\theta
\end{equation}
and
\begin{equation} \label{Zphi2}
  \lim_{\delta\to 0} \delta^{-1}\mu_0(I) =
  \int \cos(\phi-\theta)\chi(\theta)\, d\theta = Z_{\phi},
\end{equation}
recall eq.\eqref{Zphi}. Therefore
$$
   \bar{\tau}_\phi = \mu_0^{\phi}(\tau_0)
$$
is the conditional expectation of $\tau$ defined in
Section~\ref{s:angular}.

Now it is easy to see that $\cos(\phi-\theta)\chi(\theta)\, d\theta$ is
the length element in the direction orthogonal to the outgoing velocity
vector (i.e., in the direction $\phi+\pi/2$). Therefore the integral in
eq.\eqref{area} is equal to the area of the billiard table, which is
$1-\pi(r_1^2+r_2^2)$ in our case. Thus
$$
  \bar{\tau}_\phi = \mu_0^{\phi}(\tau_0) =
\frac{1-\pi(r_1^2+r_2^2)}{2(r_1+r_2)},
$$
which is constant (independent of $\phi$).

The above argument generalizes to any Sinai billiard with convex
obstacles $B_1,\ldots,B_p$. For each obstacle $B_k$ and angle $\phi$ we
denote by ${\rm width}_{\phi}(B_k)$ the ``width'' of $B_k$ in the
direction orthogonal to $\phi$, i.e., the length of the projection of
$B_k$ onto a line orthogonal to all velocities running at the angle
$\phi$. Then by the above argument we have
\begin{equation} \label{angle}
   \bar{\tau}_{\phi} =
   \frac{{\rm Area}(Q)}{\sum_k {\rm width}_{\phi}(B_k)}.
\end{equation}
This formula holds for each $\phi$ except directions in which billiard
trajectories can run indefinitely without collisions. If the horizon is
finite, no such trajectory exists, and eq.\eqref{angle} holds for
every $\phi$. If the obstacles are circular disks, as they are in our
studies, the ``width'' of $B_k$ is just its diameter, and comparing
eq.\eqref{angle} with eq.\eqref{bartau} we see that $\bar{\tau}_{\phi}$
is constant, i.e., independent of $\phi$.

It is not hard to see that averaging $\bar{\tau}_\phi$ over $\phi$
gives
\[
   \frac{1}{2\pi}\,
   \int_{-\pi}^{\pi} \bar{\tau}_\phi \, d\phi = \bar{\tau}_0 = \mu_0(\tau_0),
\]
the classical (unconditional) mean free path, which is known to be
\begin{equation} \label{bartau}
   \bar{\tau}_0 = \frac{\pi\cdot {\rm Area}(Q)}{{\rm length}(\partial Q)},
\end{equation}
see \cite[Section 2.13]{CM}. If $\bar{\tau}_\phi$ is constant, then of
course $\bar{\tau}_\phi = \bar{\tau}_0$ for all $\phi$.

\end{appendices}

\end{document}